# Interactions and Actions in One Touch Gesture Mobile Games

Misbahu S. Zubair
University of Central Lancashire
Preston, UK
mszubair@uclan.ac.uk

Salim Muhammad
Yusuf Maitama Sule University
Kano, Nigeria
muhdsalim.a@gmail.com

## ABSTRACT

A player plays a game by sending messages into the game world using an interaction technique. These messages are then translated into actions performed on or by game object(s) towards achieving the game's objectives. A game's interaction model is the bridge between the player's interaction and its in-game actions by defining what the player may and may not act upon at any given moment. This makes the choice of interaction technique, its associated action(s), and interaction model critical for designing games that are engaging, immersive, and intuitive to play. This paper presents a study focused on One-Touch-Gesture (1TG) mobile games, with the aim of identifying the touch gestures used in popular games of this type, the types of in-game actions associated with these gestures, and the interaction models used by these games. The study was conducted by reviewing 77 of the most popular games in the last two years through playtesting by two researchers. The results of the study contribute to existing knowledge by providing an insight into the interactions and actions of popular 1TG games and providing a guide to aid in developing games of the type.

## 1 Introduction

A player plays a game by sending messages into the game world using an interaction technique, these messages are then translated into actions performed on or by game object(s) towards achieving the game's objectives [2,11]. A game's interaction model is the bridge between the player's interaction and the in-game actions by defining what the player may and may not act upon at any given moment [2]. This makes the choice of interaction technique, its associated action(s), and interaction model critical for designing games that are engaging, immersive, and intuitive to play [5,6,26]. Casual games played on smartphones and tablets are mostly designed to be interacted with using touch gestures. The types of casual games that have the highest share of mobile game players (the time fillers [25]) are those with low control complexity i.e. played with one or two simple touch gestures [9]. Despite the demand for these types of games, no research was found in the literature on their gestures, actions and interaction models. Therefore there is a need to study the touch gestures used in these games, but also how they are linked to in-game actions [15].

This paper presents a study focused on One-Touch-Gesture (1TG) games i.e. games played with only one gesture, with the aim of identifying the touch gestures used in popular 1TG games, the type of in-game actions associated with these gestures, and the interaction models used by these games. The results of the study contribute to existing knowledge by providing insight into the interactions and actions of popular 1TG games and providing a guide to aid in developing games of the type.

## 2 Literature Review

The introduction of smartphones with touch screen capabilities, multi-touch support and their resulting mass adoption [5] led to the acceptance of touch gestures as the new dominant interaction techniques for mobile devices, even though these devices have sensors that can for other interaction techniques [5,10,12,30]. Studies have investigated the use of touch gestures for specific target groups such as children (e.g. [1,23,24]) and the elderly (e.g. [7,32]), but only one study was found in the literature on the use of touch gestures in popular games [34]. However, the study's aim was not to identify popular touch gestures but to identify similar gameplay.

Some of the most common touch gestures used by mobile games and other applications as reported in the literature include, but are not limited to, tap[1,20,24,31,33], swipe[33], drag[1,20,24,31,33], slide[1], double-tap[24], long pressed[24], press [20,33], pinch[1,31], flick[1,31], hold [13]. Tap is the simplest gesture that can be performed, and it involves touching the surface of the screen with a finger and then quickly lifting the finger [20,29,31,33]. The double-tap gesture is just a quick double execution of the tap gesture [29]. The pinch gesture is a multi-touch gesture i.e. requires the use of two fingers to mimic a pinch while in contact with the screen [29,31]. Swipe[33] and flick[1,31] both refer to the same gesture, a unidirectional touch gesture that begins with contact with the screen using a finger and then quick movement of that finger in a given direction [14,31]. Similarly long pressed [24], press [20,33] and hold [13] all refer to the same gesture performed by touching the surface of the screen and then staying in contact without moving the finger. Lastly, drag and slide were referred to as a single touch gesture by Aziz et al. [1], however, each of these gestures is unique. Drag requires the user to touch the screen and then move their finger over the screen in any direction or pattern without losing contact [19,20]. Although a slide also requires the user to make contact then move their finger over the screen, the movement performed in a slide gesture is unidimensional [29]. The key difference between swipe and slide is that swipe is used to input direction (e.g. swipe right or swipe right) while slide is used to input direction and distance (e.g. sliding a volume knob to adjust sound).

When a player interacts with a game using a touch gesture or any other interaction technique, an action in the game world is performed in response to that interaction [2,11]. Actions are "the verb of the game and the way in which the player usually thinks about his play" [2]. Galloway [11] categorizes a players actions as diegetic (occur within the world of the gameplay) or non-diegetic (occurring outside gameplay) actions. This categorization only differentiates between actions that affect the game's world and those that do not, it does not take into consideration the control players have over actions that affect the game world. This is important due to the varying levels of complexity of in-game actions. They can be as simple as taking a step, or as complex as negotiating the sale of a property.

A game's interaction model determines how players interact with a game by defining what players may and may not act upon at any given moment [2]. It is the bridge between players and game worlds i.e. it translates players' interactions into actions within the game world. In addition to enabling interactions, it can also be used to automate certain actions so that the player does not have to perform them, thus reducing complexity and improving accessibility [35]. Common interaction models include multipresent, avatar-based, party-based, and contestant models. Multipresent model allows players to interact with different parts of the game world whenever they want to i.e., they can perform actions anywhere within the game world. Avatar-based, on the other hand, provides players with a representative in the game world in the form of a character and players are only able to interact with the game world through the actions of that character. In party-based model, the player controls a group of characters. Finally, the contestant model lets the player take actions as if the player is a contestant on a TV show i.e. by answering questions and making decisions. Despite the importance of interaction models and the effect they have on gameplay, no work was found in the literature on their use in popular mobile games.

This review of the literature shows there is no existing work that analyses popular mobile games, in general, to identify their touch gestures, categorize their in-game actions and identify associations with touch gestures; and identify their interaction models.

## 3 Method

The data analyzed to achieve the aim of this research was collected together with data for achieving other research objectives. Therefore, this section describes the data collection process using examples relevant to the data analyzed in this paper. App Annie's monthly rankings for the top 10 most downloaded mobile games worldwide for iOS and Google Play combined were used to identify the most downloaded games for every month from July 2018 to June 2020 [3]. Most related research studies were found to use either App Store (iOS)[1,8] or Google Play[34] games, and some of these studies used rankings restricted to specific geographical areas[4]. This study's approach was taken to identify games that transcend specific platforms, geographical regions, and to have access to historical data from both App Store and Google Play in a single location [17]. 77 unique titles were identified due to having multiple games appearing in more than one month's most downloaded games.

The identified games were reviewed through play-testing by two researchers (R1 and R2), the authors. R1 downloaded the identified games from the App Store and play-tested them on an iOS smartphone while R2 downloaded the games from the Google Play Store and play-tested them on an android smartphone. Play-testing by the authors was also the methodology used by Legner et al. [21], although they only tested on one smartphone platform. Each researcher played at least 10 levels or matches of each game (where possible) and recorded observations about various features including its interaction model(s), its touch gestures (if any) and associated action(s). Touch gesture identified in the literature review and interaction models defined by [2] were assigned (where possible) based on observations during play-testing. Actions observed to be performed by each touch gesture were provided descriptive names by each researcher before a final standard name was agreed by both researchers. Actions performed once, or in iterations were recorded using the present tense of the actions' verb (e.g cut), while actions performed continuously were recorded using the present continuous form of the actions' verb (e.g moving). In cases where extremely challenging games were play-tested (e.g. Granny), additional was data was sought through other means e.g. information provided on application stores and game streams on YouTube [22].

To ensure consistency and accuracy of the observations made by both researchers, the researchers met at the end of every day to discuss their observations, replay games where contradicting observations were made, and then record a final version of the data for the game. Two games (Merge Planes and Ink Inc) were only play-tested on iOS as they were removed from Google Play store during data collection.

## 4 Data Analysis and Results

58% of the games reviewed were found to be 1TG mobile games. Five touch gestures were found to be used by these games, they are: drag, hold, tap, slide and swipe. The gestures and the number of games that utilize them are shown in figure 1.

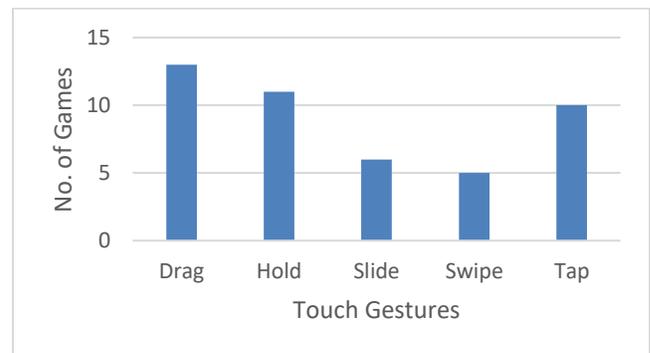

Figure 1: Touch gestures and the number of games they are used in.

To categorize in-game actions, an approach that considers the control a player has over the execution of an action was created. The categorization divides actions into two groups (initiated and controlled actions). Initiated actions are initiated by a user's interaction and then executed either instantly (e.g. cutting a rope in Rescue Cut) or over several frames (e.g. slapping an opponent in Slap Kings) until they are completed or cancelled. Controlled actions, on the other hand, require continuous interaction from the user from the moment of initiation to keep executing i.e. they are performed as long as the user continues the interaction that initiated the action and are stopped as soon as the user stops that interaction (e.g. holding to keep swinging in Stickman Hook). Controlled actions give the user the ability to determine when to start the action, and how long the action lasts.

All actions identified during data collection were categorized into these two categories. Since the aim of the study is to identify action types not unique actions, meaningful names were used to represent similar actions e.g. moving around (Rise Up) and moving downwards (ASMR Slicing) were represented as moving. Table 1 shows the grouping of the identified actions and their type, while table 2 shows the identified gestures, the type and example of actions they can be used to can perform

**Table 1: Categorization of recorded actions.**

| Action Type | Recorded Actions |
|---|---|
| Controlled | Aiming, Breaking, Digging, Drawing, Driving, Moving, Obstacle Racing, Opening and Closing, Parking, Reshaping, Rotating, Shooting, Steering, Swinging |
| Initiated | Cut, Jump Flip, Move, Point, Pull, Roll, Select, Shoot, Slap |

**Table 2: Gestures, their associated action types and example.**

| Gesture | Action Type | Action Example |
|---|---|---|
| Drag | Controlled | Drawing |
| Hold | Controlled | Obstacle Racing |
| Slide | Controlled | Steering |
| Swipe | Initiated | Move |
| Tap | Initiated | Slap |

During data collection, two games were found to require the player to interact with the game world as if interacting with a single game object. One of the games is Helix Jump, in which the player controls the world (a helix maze) so that a ball can find its way to the bottom of the maze. Therefore, world-based interaction model was created to accommodate these games.

Figure 2 shows the interaction models identified, the number of games using them, and the touch gestures used by these games.

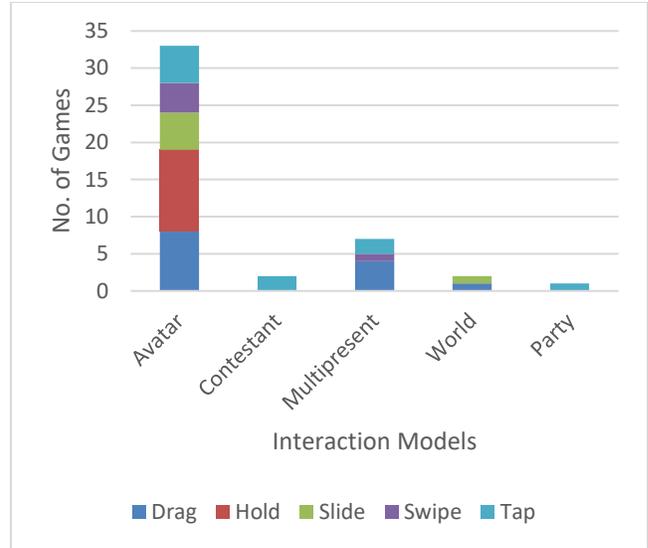

**Figure 2:** Interaction models, the number games in which they were found to be used, and the gestures used by the games.

## 5 Discussion

To discuss the results in detail, this section is divided into the following sub-sections: Gestures, Actions, Interaction Models and Automation; Ease of Avatar Control; and Interaction Models, Creativity and User Experience.

### 5.1 Gestures, Actions, Interaction Models and Automation

This results showed five gestures used by popular 1TG mobile games, they are drag, hold, slide, swipe and tap. Multi-touch gestures such as pinch were not found to be used by popular 1TG games, this is likely due to difficulties users associate with performing them compared to single-touch gestures [16,27].

The top two most used touch gestures by the games play-tested are drag and hold, and they were both used for performing controlled actions. Slide, another gesture used for performing controlled actions was the fourth most popular gesture recorded. The high occurrence of gestures used for controlled actions is as a result of their use for movement actions in games with avatar-based interaction model [2] (the highest occurring interaction model). However, each was used for a different type of movement. Drag was used for multidimensional movement (e.g. in games Popular Wars) where the avatar can be moved in all directions, vertically and horizontally, in the game world. Slide was used for movement along a single dimension and was mostly used in situations where the avatar moves automatically in one direction (e.g. vertically), and the player is charged with moving the avatar horizontally (within the constraints of the screen) (e.g. in Clean Road). Lastly, hold was used for movements that have predefined and automated actions associated with them. The avatar moves as these actions are performed as long as the player continues their hold gesture (e.g. obstacle racing in Epic Race).

In general, hold was used to perform controlled actions with the most automation. The challenge in the resulting game then becomes that of timing i.e. the player must hold for just the right amount of time for an action to complete, releasing the hold too early or too late can lead to failure in the game. For example, the game Drive and Park turns parking a car into a timing challenge by having the player hold to perform a parkin action. Releasing too early leaves the car partially on the road, releasing too late gets the car parked on the pavement.

Gestures for initiated actions, swipe and tap, were also used for automated movement actions in avatar-based interaction games. Swipe was used for providing a direction for an avatar to move towards (e.g. House Paint). Tap, on the other hand, was used in Hunter Assassin to point to a location or target for the avatar to automatically moves to or to attack.

Multipresent interaction model was found to have the second-highest number of games. The 1TG mobile games that used this model only used three gestures: drag, swipe and tap. As in avatar-based games, drag was used for movement in these games, but it was more commonly used for drawing (e.g. in Love Balls). Swipe was used in only one game, Rescue Cuts. Finally, tap was used in two multipresent games: to point at clues in Find Difference: Detective; and to initiate the pulling of a pin in Pull the Pin.

All the 1TG games with party-based and contestant-based interaction models only used the tap gesture. Tap was used to select the right answer/option in contestant-based 1TG games; to roll the dice and initiate player movement in Ludo Kings, the only party-based 1TG game found.

Finally, the two world-based 1TG games used gestures for controlled actions. Drag was used in Polysphere to rotate the world in any dimension, while slide was used to rotate the world along the x-axis in Helix Jump.

Based on the discussion above, drag should be used to control multidimensional actions (e.g. moving, dragging, drawing), slide to control unidimensional actions (e.g. increasing and reducing, opening and closing), and hold to control actions that are automated and associated with a timing challenge (hold until action is completed). Swipe should be used to initiate actions that require a direction as input (e.g. turn right, jump up, shoot forward), and tap should be used to initiate actions that are performed instantly or those that are fully automated.

### 5.2 Ease of Avatar Control

From the discussion above, it can be seen that avatar movement is the most performed action by the identified touch gestures. All the gestures for performing avatar movement were found to not require direct interaction with the avatar and can be performed anywhere on the screen. This approach addresses issues associated with touch screens such as difficulty performing gestures while distracted or doing other activities [28], allows players to perform gestures on the screen near where they are holding the device which is where researchers found most gestures to be performed [33], thus preventing issues like blocking the avatar with the hand while playing [5].

Finally, although tap and hold gestures allowed the player to perform more actions with fewer interactions through automation, drag, swipe and slide were found to be more consistent with real-world behaviours which made them more enjoyable [18]

### 5.3 Interaction Models, Creativity, and User Experience

Several games were reviewed during data collection that allowed players to draw and summon out of nowhere a game object of their liking to help them achieve their objective. This interaction approach creates countless possibilities for what a player can achieve and how it can be achieved and can be used in serious games for teaching creativity and problem-solving.

Another interesting and creative observation made by the authors relates to two games, Helix Jump and Stack Balls. These two games will look very similar to the spectator, as they both require the player to get a bouncing ball to the bottom of a helix platform. However, to the player, these games can be very different as Helix Jump uses world-based interaction model and gives the player control of the helix as the ball bounces automatically, while Stack Balls uses avatar-based interaction model and gives the player control of the ball while the helix rotates automatically. This shows how the choice of interaction model can be used creatively in creating mobile games with unique experiences.

## 6  Conclusion

This study aimed to identify the gestures used by 1TG mobile games, the types of actions they are associated with, and the interaction models used by these games. Five touch gestures (drag, hold, tap, slide and swipe) were found to be used by 45 1TG mobile games that were amongst the most downloaded games in the past two years. The actions performed using these gestures were categorized into two groups: controlled actions (performed using drag, hold or slide) are performed as long as the player continues their touch interaction (e.g. dragging to move a game object); and initiated actions (performed using tap or swipe) are initiated by the player and are either performed instantly or over several frames automatically until completed or cancelled. Avatar-based interaction model was found to be the most popular interaction model used by the 1TGs, followed by multipresent, contestant-based, world-based, and party. World-based interaction model was created to accommodate 1TGs that allow the player to interact with the whole game world as if it was a single game object.

The results of this study contribute to existing knowledge by identifying the touch gestures used in popular 1TG games, providing a categorization for in-game actions, associating actions types with touch gestures, and identifying interaction models used by 1TG games. The study also contributes guidance for choosing the right combination of touch gesture and in-game actions and encourages creative use of interaction models for more engaging and immersive games mobile games.

As future work, we will aim to study how player experience is affected by the interaction model used by a game.